\documentclass[10pt]{article}
\usepackage{amsmath}
\usepackage{amssymb}
\usepackage{natbib}

\begin{document}
\bibliographystyle{plainnat}
\setcitestyle{numbers,square}

\title{Gravitational quantization of exoplanet orbits in                  \\
                   HD~10180, Kepler-32, Kepler-33, Kepler-102,            \\
                   and Kepler-186}

\author{Vassilis S. Geroyannis                                            \\
        Department of Physics, University of Patras, Greece               \\  
        vgeroyan@upatras.gr}

\maketitle

\begin{abstract}
The so-called ``global polytropic model'' is applied to the numerical study of the exoplanet systems HD~10180, Kepler-32, Kepler-33,  Kepler-102, and Kepler-186. We compare computed distances of planets from their host stars with corresponding observations and discuss some further orbit predictions made by the model.   
\end{abstract}

\section{Introduction}
\label{intro}
This work is continuation of two previous papers regarding exoplanet systems (\citep{G14}, \citep{G15}). We do not intend to repeat here issues developed in these papers; interested readers can find a detailed account of the so-called ``global polytropic model''  --- which assumes hydrostatic equilibrium for a planetary system and proceeds to relevant computations --- in \citep{G14} (Secs.~2, 3, and references therein). Here, we study numerically the exoplanet systems HD~10180, Kepler-32, Kepler-33, Kepler-102, and Kepler-186. 

In the following tables, the first root $\xi_1$ of the Lane--Emden function $\theta$, coinciding with the radius of the host star, is expressed in both ``classical polytropic units'' (cpu) --- in such units, the length unit is equal to the polytropic parameter $\alpha$ (\citep{GVD14}, Eq.~(3b)) --- and solar radii $R_\odot$. All other orbit radii are expressed in AU.

\section{The HD~10180 System}
For the HD~10180 system (\citep{LSM10}, \citep{T12}, \citep{KG14}; evidence for 9 planets in this system is discussed in \citep{T12}), our results are given in Table~\ref{hd10180}. The minimum sum of absolute percent errors is  
\begin{equation}
\begin{aligned}
\Delta_\mathrm{min} 
\biggl( n_\mathrm{opt}(\mathrm{HD\,10180}) & =
        3.060; \, q_\mathrm{b} = 2, \, q_\mathrm{c} = 3, \biggr. \\
      & \biggl.   q_\mathrm{d} = 4, \, q_\mathrm{e} = 5, \, q_\mathrm{f} = 6, \,
                  q_\mathrm{g} = 8, \, q_\mathrm{h} = 11 
\biggr) \simeq 83.4.
\label{DminJnow}
\end{aligned}
\end{equation}
Smaller error is that for f's distance, $\simeq 0.05\%$, and larger one is that for b's distance, $\simeq 43.6\%$. The average error for the computed orbit radii of the 7 planets in HD~10180 is $\simeq 11.9\%$.

Regarding the large error involved in the distance of b (the larger one among the systems examined here), it may arise due to the proximity of the shell No~2, accomodating the planet b, with the host star. In fact, HD~10180 is the only system examined here with a planet hosted in the innermost shell. This situation is similar to that revealed for the planet e of the 55 Cnc system (\citep{G14}, Sec.~4 and Table~1) with its computed distance differing $\sim 32\%$ from the observed one. Alternatively, an interesting conjecture --- made firstly for the planet f of the  HD~40307 system (\citep{G15}, Eq.~(2) and Sec.~3.1) --- is to associate this distance with the right average-density orbit $\alpha_\mathrm{R2} = 0.0234$~AU, provided that the maximum-density orbit $\alpha_{2}$ is already occupied by another planet not yet observed. If so, then the error for b's distance drops to $\simeq 5.4\%$, the minimum sum of absolute percent errors drops to $\simeq 45.2\%$, and the average error for the HD~10187 system drops to $\simeq 6.5\%$. 

For convenience, we will use hereafter the abbreviations LADC and RADC for the ``left average-density orbit conjecture'' and the ``right average-density orbit conjecture'', respectively. 

Next, regarding the evidence given in Table~3 of \citep{T12} for the existence of two more planets i and j (9-planet solution for the system HD~10180) at distances 0.09~AU and 0.33~AU from the host star, we find that the distance of i can be associated with the left average-density orbit $\alpha_\mathrm{L4} = 0.1040$~AU, so differing $\simeq 15.6\%$ from that. The distance of j can be associated in turn with the right average-density orbit $\alpha_\mathrm{R5} = 0.3292$~AU, thus differing $\simeq 0.2\%$ from that. In both shells No~4 and No~5, the maximum-density orbits $\alpha_4$ and $\alpha_5$ are occupied by the already observed planets d and e. Thus the condition: ``provided that the maximum-density orbit is already occupied by another planet'' in our conjecture becomes a fact for these shells.

\section{The Kepler-32 System}
For the Kepler-32 system (\citep{SAM12}; observational data used in the comparisons are from this paper), the optimum case found by the global polytropic model (Table~\ref{k32}) has minimum sum of absolute percent errors   
\begin{equation}
\begin{aligned}
\Delta_\mathrm{min} 
\biggl( n_\mathrm{opt}(\mathrm{Kepler-32}) & =
        2.608; \, q_\mathrm{f} = 3, \, q_\mathrm{e} = 5, \biggr. \\
      & \biggl.   q_\mathrm{b} = 5, \, q_\mathrm{c} = 6, \, q_\mathrm{d} = 8 
\biggr) \simeq 45.2.
\label{DminS}
\end{aligned}
\end{equation}
Smaller error is that for f's distance, $\simeq 2.2\%$; while larger one is that for b's distance, $\simeq 23.9\%$. The average error for the computed distances of the 5 planets in Kepler-32 is $\simeq 9.1\%$.

Regarding the large error involved in b's distance, it seems interesting to invoke for this case both LADC and RADC. In particular, provided that the maximum-density orbit $\alpha_5 = 0.0395$~AU is already occupied by a planet not yet observed, the planet e is hosted on the left average-density orbit $\alpha_{\mathrm{L}5} = 0.0337$~AU (as in Table~\ref{hd10180}), and the planet b occupies the right average-density orbit $\alpha_{\mathrm{R}5} = 0.0457$~AU. Thus we need to employ all three available hosting orbits within the shell No~5; a similar situation has been revealed in the discussion on certain orbit predictions regarding the Kepler-275 system (\citep{G15}, Sec.~3.5). Accordingly, the difference for b's distance drops to $\simeq 11.9\%$, the minimum sum of absolute percent errors drops to $\simeq 33.2$, and the average error for the computed distances of the planets in Kepler-32 drops to $\simeq 6.6\%$.

\section{The Kepler-33 System}
Concerning the Kepler-33 system (\citep{LMR12}; observational data used in the comparisons are from this paper), the optimum case computed by the global polytropic model (Table~\ref{k33}) gives minimum sum of absolute percent errors 
\begin{equation}
\begin{aligned}
\Delta_\mathrm{min} 
\biggl( n_\mathrm{opt}(\mathrm{Kepler-33}) & =
        2.592; \, q_\mathrm{b} = 4, \, q_\mathrm{c} = 5, \biggr. \\
      & \biggl.   q_\mathrm{d} = 6, \, q_\mathrm{e} = 6, \, q_\mathrm{f} = 7 
\biggr) \simeq 30.8.
\label{DminS}
\end{aligned}
\end{equation}
Smaller error is that for e's distance, $\simeq 0.2\%$, while larger error is that for d's distance, $\simeq 15.7\%$. The average error for the computed distances of the 5 planets in Kepler-33 is $\simeq 6.2\%$.

\section{The Kepler-102 System}
The optimum case for the 5-planet Kepler-102 system (Table~\ref{k102}) gives minimum sum of absolute percent errors 
\begin{equation}
\begin{aligned}
\Delta_\mathrm{min} 
\biggl( n_\mathrm{opt}(\mathrm{Kepler-102}) & =
        2.605; \, q_\mathrm{b} = 5, \, q_\mathrm{c} = 5, \biggr. \\
      & \biggl.   q_\mathrm{d} = 6, \, q_\mathrm{e} = 7, \, q_\mathrm{f} = 8 
\biggr) \simeq 23.1.
\label{DminS}
\end{aligned}
\end{equation}
Smaller error is that for b's distance, $\simeq 0.05\%$, while larger one is that for d's distance, $\simeq 12.2\%$. The average error for the computed distances of the 5 planets in Kepler-102 is $\simeq 4.6\%$.

\section{The Kepler-186 System}
The optimum case for the 5-planet Kepler-186 system (\citep{BRP14}, \citep{BRLe14}) gives minimum sum of absolute percent errors (Table~\ref{k186})
\begin{equation}
\begin{aligned}
\Delta_\mathrm{min} 
\biggl( n_\mathrm{opt}(\mathrm{Kepler-186}) & =
        2.530; \, q_\mathrm{b} = 6, \, q_\mathrm{c} = 7, \biggr. \\
      & \biggl.   q_\mathrm{d} = 8, \, q_\mathrm{e} = 9, \, q_\mathrm{f} = 15 
\biggr) \simeq 34.3.
\label{DminS}
\end{aligned}
\end{equation}
Smaller error is that for b's distance, $\simeq 1.9\%$, while larger error is that for c's distance, $\simeq 13.2\%$. The average error for the computed distances of the 5 planets in Kepler-186 is $\simeq 6.9\%$.

Regarding the assumption of an extra planet between e and f and the induced dynamical consequences on the Kepler-186 system, for this extra planet is given in \citep{BRP14} (Sec.~5.3) an orbit prediction of 0.233~AU. On the other hand, the global polytropic model gives for the shells between No~9 and No~15: $\alpha_{10} = 0.1419$~AU, $\alpha_{11} = 0.1887$~AU, $\alpha_{12} = 0.2496$~AU, $\alpha_{13} = 0.2518$~AU, and $\alpha_{14} = 0.3277$~AU. Thus $\alpha_{12}$ is most likely the orbit hosting the assumed extra planet, having a difference $\sim 7\%$ from the distance predicted in \citep{BRP14}.

\begin{table}
\begin{center}
\caption{The HD~10180 system: central body $S_1$, i.e. the host star HD~10180, and polytropic spherical shells of the planets b, c, d, e, f, g. For successive shells $S_j$ and $S_{j+1}$, inner radius of $S_{j+1}$ is the outer radius of $S_j$. All radii are expressed in AU, except for the host's radius $\xi_1$. Percent errors $\%E_j$ in the computed orbit radii $\alpha_j$ are given with respect to the corresponding observed radii $A_j$, $\%E_j = 100 \times |(A_j - \alpha_j)| / A_j$. Parenthesized signed integers following numerical values denote powers of 10. \label{hd10180}}
\begin{tabular}{lrrl} 
\hline \hline
Host star HD~10180 -- Shell No                  & 1              \\
$n_\mathrm{opt}$                                & $3.060\,\,\,(+00)$  \\
$\xi_1$ (cpu)                                   & $7.1385(+00)$  \\
$\xi_1$ ($R_\odot$)                             & $1.20 \ \ \ (+00)$  \\
\hline
            & & $A$~~~~~~ & $~~~~\%E$ \\  
\hline

b -- Shell No                                 & 2                 \\
Inner radius, $\, \xi_1$                      & $5.5827(-03)$     \\              
Outer radius, $\xi_2$                         & $3.0524(-02)$     \\
Orbit radius, $\, \alpha_\mathrm{b}=\alpha_2$ & $1.2513(-02)$ & $2.22(-02)$ & $4.36(+01)$ \\
\hline

c -- Shell No                                 & 3                 \\
Outer radius, $\xi_3$                         & $9.0232(-02)$     \\
Orbit radius, $\, \alpha_\mathrm{c}=\alpha_3$ & $5.2398(-02)$ & $6.41(-02)$ & $1.83(+01)$ \\
\hline

d -- Shell No                                 & 4                 \\
Outer radius, $\xi_4$                         & $2.0074(-01)$     \\
Orbit radius, $\, \alpha_\mathrm{d}=\alpha_4$ & $1.3373(-01)$ & $1.286(-01)$ & $3.99(+00)$ \\   
\hline

e -- Shell No                                 & 5                 \\
Outer radius, $\xi_5$                         & $3.7905(-01)$     \\
Orbit radius, $\, \alpha_\mathrm{e}=\alpha_5$ & $2.7529(-01)$ & $2.699(-01)$ & $1.99(+00)$               \\ 
\hline

f -- Shell No                                    & 6                \\
Outer radius, $\xi_6$                            & $6.4201(-01)$     \\
Orbit radius, $\, \alpha_\mathrm{f}=\alpha_6$    & $4.9270(-01)$ & $4.929(-01)$ & $4.08(-02)$               \\   
\hline

g -- Shell No                                    & 8                \\
Inner radius, $\, \xi_7$                         & $1.0079(+00)$     \\
Outer radius, $\xi_8$                            & $1.4938(+00)$     \\
Orbit radius, $\, \alpha_\mathrm{g}=\alpha_8$    & $1.2255(+00)$ & $1.422(+00)$ & $1.38(+01)$               \\   
\hline

h -- Shell No                                    & 11                \\
Inner radius, $\, \xi_{10}$                      & $2.9012(+00)$     \\
Outer radius, $\xi_{11}$                         & $3.8603(+00)$     \\
Orbit radius, $\, \alpha_\mathrm{h}=\alpha_{11}$ & $3.3449(+00)$ & $3.4(+00)$ & $1.62(+00)$               \\   
\hline

\end{tabular}
\end{center}
\end{table}

\begin{table}
\begin{center}
\caption{The Kepler-32 system: central body $S_1$, i.e. the host star Kepler-32, and polytropic spherical shells of the planets f, e, b, c, d. Other details as in Table \ref{hd10180}. \label{k32}}
\begin{tabular}{lrrl} 
\hline \hline
Host star Kepler-32 -- Shell No                 & 1                   \\
$n_\mathrm{opt}$                                & $2.608\,\,\,(+00)$  \\
$\xi_1$ (cpu)                                   & $5.6307(+00)$  \\
$\xi_1$ ($R_\odot$)                             & $5.3 \ \ \ \ \, (-01)$  \\
\hline
            & & $A$~~~~~~ & $~~~~\%E$ \\  
\hline

f -- Shell No                                 & 3                 \\
Inner radius, $\, \xi_2$                      & $8.9187(-03)$     \\              
Outer radius, $\xi_3$                         & $1.6317(-02)$     \\
Orbit radius, $\, \alpha_\mathrm{f}=\alpha_3$ & $1.3290(-02)$ & $1.3(-02)$ & $2.23(+00)$ \\
\hline

e -- Shell No                                 & 5                 \\
Inner radius, $\, \xi_4$                      & $2.9566(-02)$     \\
Outer radius, $\xi_5$                         & $5.2173(-02)$     \\
Orbit radius, $\, \alpha_\mathrm{e}=\alpha_{\mathrm{L}5}$ & $3.3697(-02)$ & $3.23(-02)$ & $4.33(+00)$ \\
\hline

b -- Shell No                                   & 5                 \\
Orbit radius, $\, \alpha_\mathrm{b}=\alpha_{5}$ & $3.9499(-02)$ & $5.19(-02)$ & $2.39(+01)$ \\   
\hline

c -- Shell No                                    & 6                \\
Outer radius, $\xi_6$                            & $7.5932(-02)$     \\
Orbit radius, $\, \alpha_\mathrm{c}=\alpha_{6}$  & $6.9474(-02)$ & $6.7(-02)$ & $3.69(+00)$               \\ 
\hline

d -- Shell No                                    & 8                \\
Inner radius, $\, \xi_7$                         & $9.8619(-02)$     \\
Outer radius, $\xi_8$                            & $1.3745(-01)$     \\
Orbit radius, $\, \alpha_\mathrm{d}=\alpha_8$    & $1.1388(-01)$ & $1.28(-01)$ & $1.10(+01)$               \\   
\hline

\end{tabular}
\end{center}
\end{table}

\begin{table}
\begin{center}
\caption{The Kepler-33 system: central body $S_1$, i.e. the host star Kepler-33, and polytropic spherical shells of the planets b, c, d, e, f. Other details as in Table \ref{hd10180}. \label{k33}}
\begin{tabular}{lrrl} 
\hline \hline
Host star Kepler-33 -- Shell No                 & 1                   \\
$n_\mathrm{opt}$                                & $2.592\,\,\,(+00)$  \\
$\xi_1$ (cpu)                                   & $5.5882(+00)$  \\
$\xi_1$ ($R_\odot$)                             & $1.615 \ \, (+00)$  \\
\hline
            & & $A$~~~~~~ & $~~~~\%E$ \\  
\hline

b -- Shell No                                 & 4                 \\
Inner radius, $\, \xi_3$                      & $4.8624(-02)$     \\              
Outer radius, $\xi_4$                         & $8.8875(-02)$     \\
Orbit radius, $\, \alpha_\mathrm{b}=\alpha_4$ & $5.8352(-02)$ & $6.544(-02)$ & $1.08(+01)$ \\
\hline

c -- Shell No                                 & 5                 \\
Outer radius, $\xi_5$                         & $1.5606(-01)$     \\
Orbit radius, $\, \alpha_\mathrm{c}=\alpha_5$ & $1.1825(-01)$ & $1.1484(-01)$ & $2.97(+00)$ \\
\hline

d -- Shell No                                   & 6                 \\
Orbit radius, $\, \alpha_\mathrm{d}=\alpha_{\mathrm{L}6}$ & $1.8568(-01)$ & $1.6053(-01)$ & $1.57(+01)$ \\   
\hline

e -- Shell No                                    & 6                \\
Outer radius, $\xi_6$                            & $2.2816(-01)$     \\
Orbit radius, $\, \alpha_\mathrm{e}=\alpha_{6}$  & $2.0618(-01)$ & $2.0656(-01)$ & $1.84(-01)$               \\ 
\hline

f -- Shell No                                    & 7                \\
Outer radius, $\xi_7$                            & $2.9495(-01)$     \\
Orbit radius, $\, \alpha_\mathrm{f}=\alpha_7$    & $2.4773(-01)$ & $2.449(-01)$ & $1.15(+00)$               \\   
\hline

\end{tabular}
\end{center}
\end{table}

\begin{table}
\begin{center}
\caption{The Kepler-102 system: central body $S_1$, i.e. the host star Kepler-102, and polytropic spherical shells of the planets b, c, d, e, f. Other details as in Table \ref{hd10180}. \label{k102}}
\begin{tabular}{lrrl} 
\hline \hline
Host star Kepler-102 -- Shell No                & 1                   \\
$n_\mathrm{opt}$                                & $2.605\,\,\,(+00)$  \\
$\xi_1$ (cpu)                                   & $5.6227(+00)$  \\
$\xi_1$ ($R_\odot$)                             & $7.4 \ \ \ \ \, (-01)$  \\
\hline
            & & $A$~~~~~~ & $~~~~\%E$ \\  
\hline

b -- Shell No                                 & 5                 \\
Inner radius, $\, \xi_4$                      & $4.1175(-02)$     \\              
Outer radius, $\xi_5$                         & $7.2596(-02)$     \\
Orbit radius, $\, \alpha_\mathrm{b}=\alpha_5$ & $5.4983(-02)$ & $5.5(-02)$ & $3.05(-02)$ \\
\hline

c -- Shell No                                 & 5                 \\
Orbit radius, $\, \alpha_\mathrm{c}=\alpha_{\mathrm{R}5}$ & $6.3616(-02)$ & $6.7(-02)$ & $5.05(+00)$ \\
\hline

d -- Shell No                                   & 6                 \\
Outer radius, $\xi_6$                           & $1.0576(-01)$     \\
Orbit radius, $\, \alpha_\mathrm{d}=\alpha_6$   & $9.6527(-02)$ & $8.6(-02)$ & $1.22(+01)$ \\   
\hline

e -- Shell No                                    & 7                \\
Outer radius, $\xi_7$                            & $1.3722(-01)$     \\
Orbit radius, $\, \alpha_\mathrm{e}=\alpha_7$    & $1.1395(-01)$ & $1.16(-01)$ & $1.77(+00)$               \\ 
\hline

f -- Shell No                                    & 8                \\
Outer radius, $\xi_8$                            & $1.9100(-01)$     \\
Orbit radius, $\, \alpha_\mathrm{f}=\alpha_8$    & $1.5828(-01)$ & $1.65(-01)$ & $4.08(+00)$               \\   
\hline

\end{tabular}
\end{center}
\end{table}

\begin{table}
\begin{center}
\caption{The Kepler-186 system: central body $S_1$, i.e. the host star Kepler-186, and polytropic spherical shells of the planets b, c, d, e, f. Other details as in Table \ref{hd10180}. \label{k186}}
\begin{tabular}{lrrl} 
\hline \hline
Host star Kepler-186 -- Shell No                & 1                   \\
$n_\mathrm{opt}$                                & $2.530\,\,\,(+00)$  \\
$\xi_1$ (cpu)                                   & $5.4292(+00)$  \\
$\xi_1$ ($R_\odot$)                             & $4.7 \ \ \ \ \, (-01)$  \\
\hline
            & & $A$~~~~~~ & $~~~~\%E$ \\  
\hline

b -- Shell No                                 & 6                 \\
Inner radius, $\, \xi_5$                      & $3.7778(-02)$     \\              
Outer radius, $\xi_6$                         & $5.3777(-02)$     \\
Orbit radius, $\, \alpha_\mathrm{b}=\alpha_6$ & $3.8502(-02)$ & $3.78(-02)$ & $1.86(+00)$ \\
\hline

c -- Shell No                                 & 7                 \\
Outer radius, $\xi_7$                         & $7.9508(-02)$     \\
Orbit radius, $\, \alpha_\mathrm{c}=\alpha_7$ & $6.4962(-02)$ & $5.74(-02)$ & $1.32(+01)$ \\
\hline

d -- Shell No                                   & 8                 \\
Outer radius, $\xi_8$                           & $1.0746(-01)$     \\
Orbit radius, $\, \alpha_\mathrm{d}=\alpha_8$   & $9.6356(-02)$ & $8.61(-02)$ & $1.19(+01)$ \\   
\hline

e -- Shell No                                    & 9                \\
Outer radius, $\xi_9$                            & $1.3021(-01)$     \\
Orbit radius, $\, \alpha_\mathrm{e}=\alpha_9$    & $1.1837(-01)$ & $1.216(-01)$ & $2.66(+00)$               \\ 
\hline

f -- Shell No                                    & 15                \\
Inner radius, $\, \xi_{14}$                      & $3.5064(-01)$     \\    
Outer radius, $\xi_{15}$                         & $3.9583(-01)$     \\
Orbit radius, $\, \alpha_\mathrm{f}=\alpha_{15}$ & $3.7423(-01)$ & $3.926(-01)$ & $4.68(+00)$               \\   
\hline

\end{tabular}
\end{center}
\end{table}

\end{document}